# A Wind Tunnel in Your Classroom: The Design and Implementation of a Portable Wind Tunnel for Use in the Science Classroom

*John H. Huckans[a,e], Nathan A. Kurz[b], Dean C. Walker[b], Carla Zembal-Saul[c] and Milton W. Cole[a,d]*, Penn State University, University Park, PA 16802

*Kimber H. Mitchell and Diane S. Reed*, Radio Park Elementary School, State College, PA 16803

## Abstract

This article describes a wind tunnel's ease of construction and its uses as a visualization tool that integrates concepts at the primary and secondary levels. An Appendix contains detailed building instructions. A web site (under construction) will provide sample lesson plans, visualization and application tips, animations and links to background information.

One hundred years ago, in the fall of 1902, Wilbur and Orville Wright tested their most successful glider at the Kill Devil Hills camp, near Kitty Hawk, North Carolina. This was not an impromptu event; it followed more than a year of testing wing sections in their homemade wind tunnel, back in Dayton, Ohio. The glider's performance confirmed calculations based on the wind tunnel data. Their ultimately successful venture provides a useful model for science teachers, students and researchers to emulate.

During the spring of 2001, a group of State College Area School District and Penn State University faculty (along with a volunteer pilot from US Airways) met to discuss ways to enhance instruction in a third and fourth grade unit, called Air and Aviation. A specific goal of these meetings was to find classroom strategies that would help students to visualize abstract concepts of flight, such as lift and drag. One idea seemed feasible: to create a small, inexpensive device that would illustrate Bernoulli's principle in a way that third graders could understand. The device began as a small wind tunnel, similar to that developed by the Wright brothers, which would allow "correctly" shaped paper airfoils (plane wings) to fly. That same device ended up being an efficient teaching tool with numerous applications.

Potential uses of our wind tunnel device (nicknamed the "Bernoulli box" and seen in figure 1) range from discovery and guided learning of qualitative aspects of lift to the possibility of adding instrumentation for quantitative studies. Documentation of the device and its possible use in the classroom appear at a web site, http://scied.ed.psu.edu/~czem/aviation/.

The air travels first through a duct fan, which expels air into a sheet-metal box, and then it flows onward through a number of flexible PVC tubes into a long, clear Plexiglas tunnel. The array of tubes was designed to reduce turbulence, which disturbs the flow out of the fan. Just imagine how many lesson plans you can create with a clear box of gentle and smoothly flowing air! A wing section placed within the box creates air-flow patterns that are fundamentally identical to those near an airplane wing in flight. See figure 2. This similarity, of course, is just the reason that wind tunnels are used in airplane design. Our apparatus lends itself well to teaching concepts such as drag, lift, velocity, streamlines, friction and the Bernoulli principle. It can also be used for more integrated projects, such as creating and analyzing airfoils, using force sensors to measure drag or wind velocity, or designing aerodynamic shapes and surfaces.

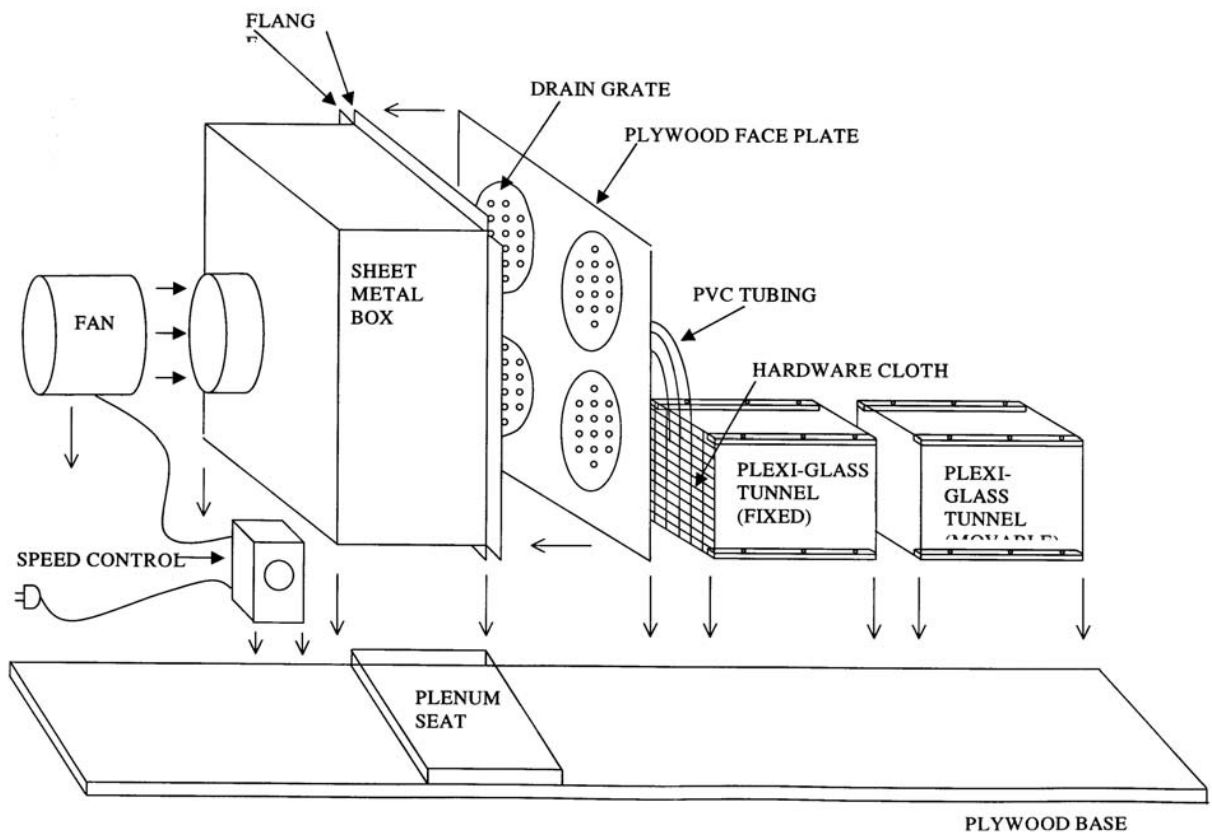

Figure 1: Bernoulli Box

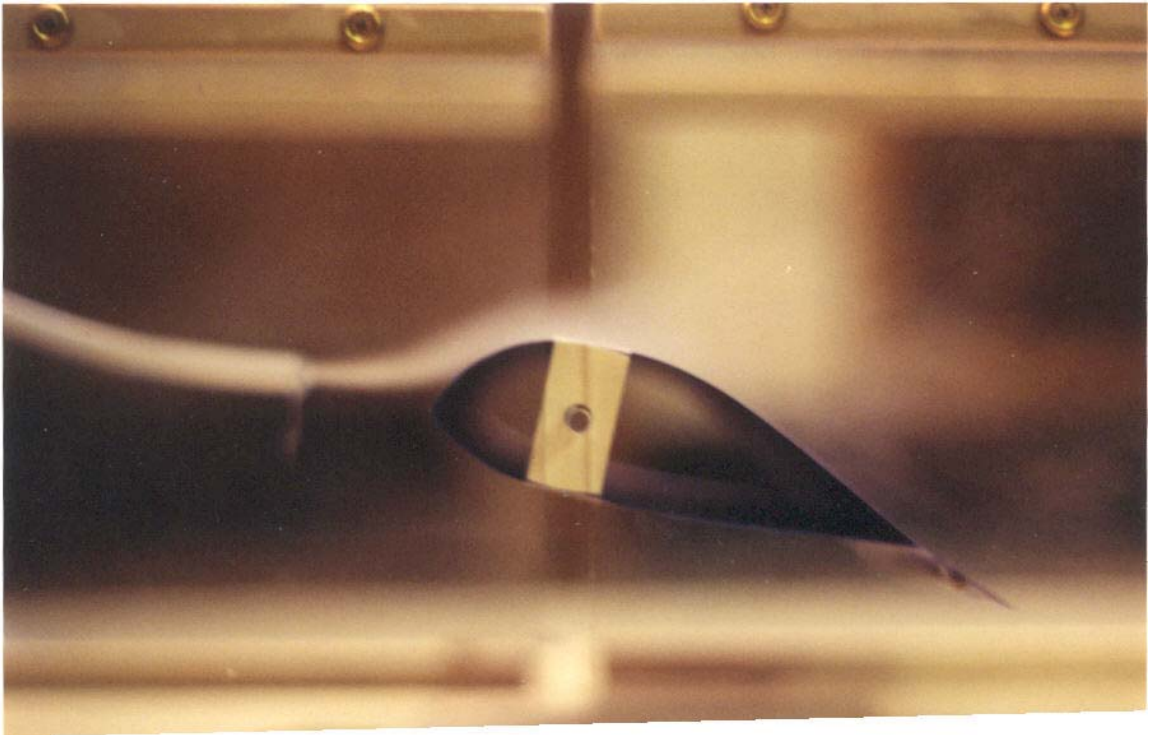

While misconceptions about the mechanics of flight abound (in people of all ages), the power of visualization that the box provides enables even young students to grasp the basic principles, in a vivid and dramatic way.  The Bernoulli box can be used in conjunction with tracers, such as dry ice mist or the cloud of a fog machine, which reveal the flow pattern of air over a wing section.  The wing sections themselves can be adjusted to manipulate their aerodynamic properties.  After presenting students with several wings of varying thickness, weight and shape, ask them "Which of these wings will fly, and why?".  Better yet- have them create their own "perfect" wing.  With this challenge, students can see how small changes in the geometry of a wing affect its behavior while it is moving through the air.  Lessons can be tailored to suit the needs of the teacher and the level of the students' prior knowledge.  A number of complete lesson plans can be found a http://scied.ed.psu.edu/~czem/aviation/.
An Appendix to this article provides detailed instructions for building your own Bernoulli box.  The website also includes a plethora of instructional tips and sources for background information so you can point the young scientists in the right direction in search of answers to their questions (about the device and the science related to it).

The Bernoulli box can be built with about $200 worth of materials and 20 hours of unskilled labor.  Teachers will realize the great investment they've made when  they see the extent to which the box allows students to develop *correct* and imaginative understanding  of such interesting, but difficult-to-

teach, topics.. Best of all, many of the concepts are covered in most states' science standards! For example, the newly adopted Pennsylvania Academic Standards for Science and Technology require that fourth grade students be able to "Identify transportation technologies of propelling, structuring, suspending, guiding, controlling and supporting."

Here is the story of the very first time the Bernoulli box was used in a classroom to teach the concept of Bernoulli's principle: The fourth grade children had been taught the terms "lift" and "drag", but most of them did not know what they mean to any significant extent. We asked them what would happen if they were to be pushed simultaneously to the left by a really strong person and to the right by a weaker person. Most answered correctly that they would start moving to the left. We then related that idea to forces on a wing, but this time the stronger push was from the bottom and the weaker push from the top. As a final preparation activity, we had the students line up along a wall and push against it as hard as they could. They could feel that they were pushing hard because they were sliding backwards on the floor. When we asked them to walk along the wall and push (similar to air molecules flowing past a surface), they were not able to push as hard. After finishing the prep work, we asked what they thought is the connection between the two activities. We could tell that most of the students had some idea but there was a need for a visual conclusion to clear up their remaining confusion. Meanwhile, they were asking about the purpose of the wind tunnel sitting in the back of the room. Figure 3 depicts the use in a classroom, guided by one of the authors (Walker).

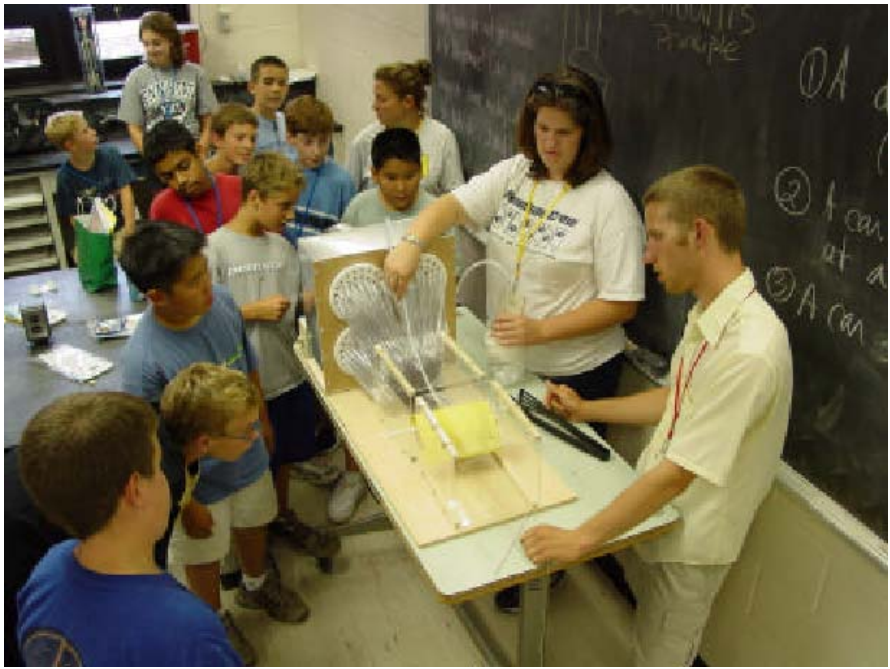

With their curiosity piqued and suspense mounting , we dropped a chunk of dry ice into the water bottle, put on its cap, and turned on the fan.  The students saw the airflow begin inside the tunnel.  The wing took off just as a real plane would, and then they saw the streamlines (figure 2).  One by one, we heard students actually exclaim "ahhh"- not the fireworks kind of "ahhh", but the "Oh, I get it" kind.  After the flow of mist stopped, we had the children record in their science journals hypothetical explanations of their observations.  They knew that as the air moved faster across a surface, it wouldn't be able to push as hard against those surfaces.  They also knew that being pushed harder from one side than the other would result in a movement in the direction of the harder push. But would they put the two ideas together?  We then had the students create wings that would test their hypotheses.  They came up in pairs to test their wings in the wind tunnel.   While many of the wings looked very similar to the original wing, practically all of the students were able to explain their designs in their own words.  In fact, some of the very oddest wing designs actually flew! Careful questioning of the designers of these unusual wings led us to believe that their creative designs were the result of a strong understanding of some concepts.  The lesson turned out to be an amazing and thrilling experience for the adults present in the classroom.  Since that lesson, the box has been used throughout the district and we have been asked to build two more to satisfy the growing audience of enthusiastic users.

We are trying to make improvements on the existing wind tunnel.  One goal is to design a less expensive version. Another potential improvement will include measuring physical data with such additions as force, velocity and pressure gauges.  While these innovations may be more appropriate for high school students, creative teachers may develop ways to use them even to stimulate elementary school students.  We have been finding through this work that the very exploration of such additions can represent lessons or entire projects in themselves.

Our research, design and construction activity was supported by the NASA Space Science Consortium at Penn State, the Eberly College of Science, the State College Area School District, the National Science Foundation-supported MRSEC at Penn State and the New York Section of the American Physical Society. We are grateful to Jayanth Banavar, Dick Brown, Dan Larson and Paul Sokol for helpful discussions and support.

[a] Department of Physics, Eberly College of Science
[b] Schreyer Honors College
[c] Department of Curriculum and Instruction, College of Education
[d] Author to whom correspondence should be addressed: mwc@psu.edu
[e] Current Address: Department of Physics and Astronomy, Univ. Maryland, College Park, 20742

# Appendix: Bernoulli Box Construction Notes

The Bernoulli box consists of a segmented plexiglass tunnel which is fed air indirectly from a high static fan.  The fan first discharges into a sheet metal box.  The box is fitted with flexible tubing which leads to the plexiglass tunnel.  The purpose of the box and tubing is to create a straight, smooth air flow within the plexiglass tunnel.

### Plywood Base and Plenum Seat

Cut a piece of 1/2 in. thick plywood into a 16 in. x 40 in. rectangle.  Cut another piece of the 1/2 in. plywood into a 8 in. x 16 in. rectangle.  The large piece is the overall base for the wind tunnel.  The smaller piece is the seat for the sheet metal box.  Glue the seat onto the base 5 in. from one end using wood glue and clamps.  Let it dry overnight.

### Sheet Metal Box

Construct a sheet metal box to receive the wind tunnel fan.  We suggest that the box be built by a local sheet metal shop.  We had our box fabricated by Duck's Sheet Metal Shop of State College, PA (814-237-3493).  They were very fast (two day turnaround), inexpensive (approximately $35 per box) and reliable.
The box should be constructed using 24 gauge galvanized sheet metal.  The dimensions are 15 in. x 15 in. x 8 in. deep.  One of the 15 in. x 15 in. ends is completely open with a 1/2 in. metal flange going all the way around.  The flange is formed by starting out with a 15 in. x 15 in. x 8.5 in. deep box, making 1/2 in. cuts at the open corners and then bending the metal back.  The other 15 in. x 15 in. end of the box is closed except for a 2 in. deep, 10 in. diameter metal inlet collar to receive the fan.  The collar is centered in the middle of the 15 in. x 15 in. box face. The remaining four smaller sides of the metal box are completely closed.  Screw the box down onto the seat using No. 8 x 1/2 in. sheet metal drill screws, letting the flange hang over the edge of the seat.  See Figure 1.

### Plywood Face Plate and Drain Grate Strainers

Cut a piece of 1/4 in. thick plywood into a 16 in. x 16 in. square.  Using a jigsaw, cut four 6 in. diameter lobed circles out of the center of the plywood as shown on Figure 2.  Obtain four 6-3/4 in. diameter polypropylene drain grate replacement strainers (PHP Item No. 30032) from Prairie Home Products of Peculiar, MO (800-367-1568). Note that you may not be able to purchase the drain grate strainers directly from PHP.  Often, manufacturers will only sell through licensed distributors or contractors with previously established accounts. We bought through the plumbing department of Lowe's in State College, PA (814-237-2100). Alternatively, you may try Home Depot (800-553-3199).

Drill eight 1/4 in. diameter holes in the lobes of the plywood face plate using the drain grates as guides in positioning the holes.  Secure the four drain grates to the plywood face plate using 1/4-20 x 1 in. machine screws with washers and nuts.  The screws should have counter-sunk style (as opposed to bugle) heads to mate with the drain grates.  Attach the face plate to the sheet metal box flanges with the drain grates facing out using No. 8 x 1/2 in. sheet metal drill screws.

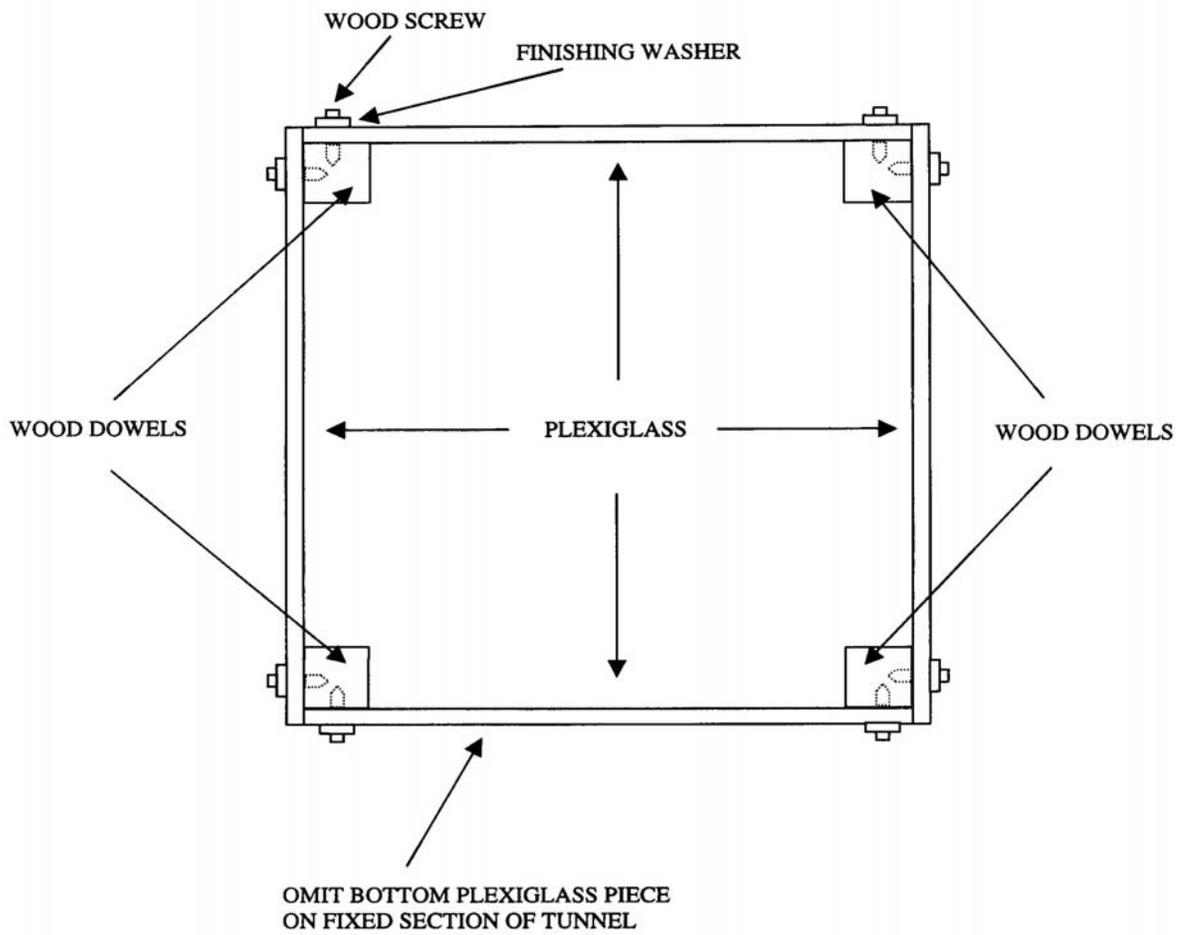

Figure 3: Plexiglass Tunnel Section

Note that one of the tunnel sections uses four plexiglass pieces.  This section is movable on the plywood base.  The other tunnel section uses only three plexiglass pieces - it has no bottom. It is screwed down on the plywood base 6 in. from the drain grates.  Both tunnel sections employ four wood dowels.  The wood dowels on the fixed tunnel section are used to secure the tunnel to the base.  We suggest using No. 6 x 1 in. wood screws and finishing washers.

Using No. 6 x 1 in. wood screws and No. 6 x 9/16 in. fender washers, attach an 8 in. x 8 in. square of hardware cloth with a 1/2 in. x 1/2 in. mesh to the end of the fixed tunnel facing the drain grates.  Hardware cloth is like heavy duty chicken wire.  It is available from most building supply stores.

The most time-consuming part of the construction and also the part requiring the most artistry is the fitting of the flexible tubing between the drain grates and the fixed tunnel.  Obtain approximately 180 ft of 7/16 in. diameter flexible PVC tubing on a roll from a local hardware or plumbing supply shop.  Make sure the fixed tunnel is screwed down tightly onto the plywood base.  Begin cutting 8 in. long tubes from the roll.  Insert one tube end into the lowest row of hardware cloth openings and the other end into the low openings of the drain grates.  Gradually work up higher, building row upon row.  The tubes will have to get longer and longer with the highest tubes approximately 11 in. long.  It is important to keep the tubes pointing straight within the tunnel.  Do not insert the fan into the sheet metal box until all the tubes are in place.  This is because you will want to get your hand inside the box to help position the tubes.

    Fan and Electronic Speed Control

The most important component of the Bernoulli box is the fan.  We suggest the model DB410 from Suncourt Inc. of Durant, IA (800-999-3267 ; www.suncourt.com).  This is a 10 in. diameter propeller-style fan with high static capability and sufficient airflow.  Insert the fan into the receiving inlet collar of the sheet metal box and fasten with the sheet metal drill screws.  Attach a safety grill at the fan inlet using sheet metal screws or construction adhesive.

Purchase a minimum 3 amp/120 VAC electronic speed control and cover plate such as by Broan-NuTone of Hartford, WI (800-558-1711 ; www.broan.com).  Also purchase a heavy duty grounded (three-prong) minimum 6 ft long power cord.  Wire the control between the power cord and the fan using standard electrical wire nuts. Install the control in an electrical box with suitable strain relief.  Again, you may not be able to purchase directly from Suncourt and Broan-NuTone.  We purchased the fans and controls from Keystone Refrigeration & Heating Supply of State College, PA (800-281-1558).  Bolt the electrical box down onto the plywood base using a No. 10 x 1 in. wood screw and 3/16 in. x 1 in. fender washer.

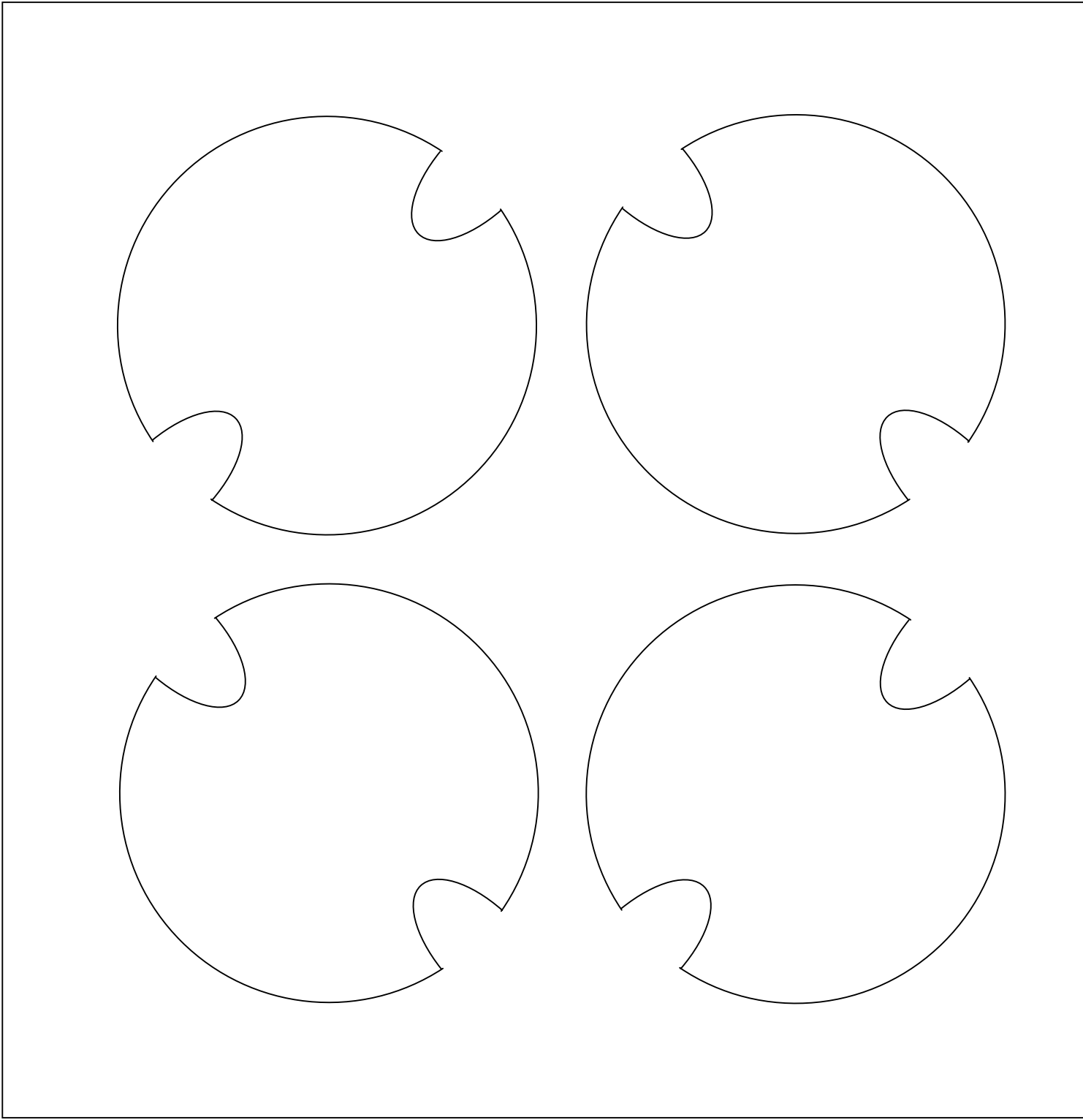

Plywood Face Plate